\documentclass{emulateapj}
\usepackage{graphicx}

\begin{document}

\shorttitle{CLASH Mass Profile of MACS J1206.2-0847}
\shortauthors{Zitrin et al.}

\slugcomment{Submitted to the Astrophysical Journal Letters}

\title{CLASH: New Multiple-Images Constraining the Inner Mass Profile of MACS J1206.2-0847}

\author{A. Zitrin\altaffilmark{1}}
\author{P. Rosati\altaffilmark{2}}
\author{M. Nonino\altaffilmark{3}}
\author{C. Grillo\altaffilmark{4}}
\author{M. Postman\altaffilmark{5}}
\author{D. Coe\altaffilmark{5}}
\author{S. Seitz\altaffilmark{6}}
\author{T. Eichner\altaffilmark{6}}
\author{T. Broadhurst\altaffilmark{7,8}}
\author{S. Jouvel\altaffilmark{9}}
\author{I. Balestra\altaffilmark{10}}
\author{A. Mercurio\altaffilmark{11}}
\author{M. Scodeggio\altaffilmark{12}}
\author{N. Ben\'itez\altaffilmark{13}}
\author{L. Bradley\altaffilmark{5}}
\author{H. Ford\altaffilmark{14}}
\author{O. Host\altaffilmark{9}}
\author{Y. Jimenez-Teja\altaffilmark{13}}
\author{A. Koekemoer\altaffilmark{5}}
\author{W. Zheng\altaffilmark{14}}
\author{M. Bartelmann\altaffilmark{15}}
\author{R. Bouwens\altaffilmark{16}}
\author{O. Czoske\altaffilmark{17}}
\author{M. Donahue\altaffilmark{18}}
\author{O. Graur\altaffilmark{1}}
\author{G. Graves\altaffilmark{19}}
\author{L. Infante\altaffilmark{20}}
\author{S. Jha\altaffilmark{21}}
\author{D. Kelson\altaffilmark{22}}
\author{O. Lahav\altaffilmark{9}}
\author{R. Lazkoz\altaffilmark{7}}
\author{D. Lemze\altaffilmark{14}}
\author{M. Lombardi\altaffilmark{23}}
\author{D. Maoz\altaffilmark{1}}
\author{C. McCully\altaffilmark{21}}
\author{E. Medezinski\altaffilmark{14}}
\author{P. Melchior\altaffilmark{24}}
\author{M. Meneghetti\altaffilmark{25}}
\author{J. Merten\altaffilmark{15}}
\author{A. Molino\altaffilmark{13}}
\author{L.A. Moustakas\altaffilmark{26}}
\author{S. Ogaz\altaffilmark{5}}
\author{B. Patel\altaffilmark{21}}
\author{E. Regoes\altaffilmark{27}}
\author{A. Riess\altaffilmark{5,14}}
\author{S. Rodney\altaffilmark{14}}
\author{K. Umetsu\altaffilmark{28}}
\author{A. Van der Wel\altaffilmark{29}}

\altaffiltext{1}{The School of Physics and Astronomy, Tel Aviv University; adiz@wise.tau.ac.il}
\altaffiltext{2}{European Southern Observatory}
\altaffiltext{3}{INAF-Osservatorio Astronomico di Trieste}
\altaffiltext{4}{Excellence Cluster Universe, Technische Universit\"at M\"unchen}
\altaffiltext{5}{Space Telescope Science Institute}
\altaffiltext{6}{Universitas Sternwarte Muenchen}
\altaffiltext{7}{University of Basque Country}
\altaffiltext{8}{IKERBASQUE, Basque Foundation for Science}
\altaffiltext{9}{University College London}
\altaffiltext{10}{MPE, Garching}
\altaffiltext{11}{INAF-Osservatorio Astronomico di Capodimonte}
\altaffiltext{12}{INAF – IASF, Milano}
\altaffiltext{13}{Instituto de Astrof\'isica de Andaluc\'ia (CSIC)}
\altaffiltext{14}{Department of Physics and Astronomy, The Johns Hopkins University}
\altaffiltext{15}{Universitat Heidelberg}
\altaffiltext{16}{University of Leiden}
\altaffiltext{17}{Institut f\"ur Astronomie der Universit\"at Wien}
\altaffiltext{18}{Michigan State University}
\altaffiltext{19}{UC Berkeley}
\altaffiltext{20}{Universidad Catolica de Chile}
\altaffiltext{21}{Rutgers University}
\altaffiltext{22}{Carnegie Institution}
\altaffiltext{23}{Dipartimento di Fisica, Universit\`a degli Studi di Milano}
\altaffiltext{24}{The Ohio State University}
\altaffiltext{25}{INAF, Osservatorio Astronomico di Bologna; INFN, Sezione di Bologna}
\altaffiltext{26}{Jet Propulsion Laboratory, California Institute of Technology}
\altaffiltext{27}{European Laboratory for Particle Physics (CERN)}
\altaffiltext{28}{Institute of Astronomy and Astrophysics, Academia Sinica}
\altaffiltext{29}{MPIA, Heidelberg}

\begin{abstract}
We present a strong-lensing analysis of the galaxy cluster MACS J1206.2-0847 ($z$=0.44) using UV, Optical, and IR, HST/ACS/WFC3 data taken as part of the CLASH multi-cycle treasury program, with VLT/VIMOS spectroscopy for some of the multiply-lensed arcs. The CLASH observations, combined with our mass-model, allow us to identify 47 new multiply-lensed images of 12 distant sources. These images, along with the previously known arc, span the redshift range $1\la z\la5.5$, and thus enable us to derive a detailed mass distribution and to accurately constrain, for the first time, the inner mass-profile of this cluster. We find an inner profile slope of $d\log \Sigma/d\log \theta\simeq -0.55\pm 0.1$ (in the range [1\arcsec, 53\arcsec], or $5\la r \la300$ kpc), as commonly found for relaxed and well-concentrated clusters. Using the many systems uncovered here we derive credible critical curves and Einstein radii for different source redshifts. For a source at $z_{s}\simeq2.5$, the critical curve encloses a large area with an effective Einstein radius of $\theta_{E}=28\pm3\arcsec$, and a projected mass of $1.34\pm0.15\times10^{14} M_{\odot}$. From the current understanding of structure formation in concordance cosmology, these values are relatively high for clusters at $z\sim0.5$, so that detailed studies of the inner mass distribution of clusters such as MACS J1206.2-0847 can provide stringent tests of the $\Lambda$CDM paradigm.

\end{abstract}


\section{Introduction}\label{intro}

Massive galaxy clusters, due to their high inner mass-density, are known to form prominent gravitational lenses. The expected distribution of lens-sizes and the abundance of giant lenses in particular, have been now established by N-body simulations \citep[e.g.,][]{Hennawi2007}, semi-analytic calculations \citep[e.g.,][]{OguriBlandford2009}, and recently, also examined observationally by a statistical analysis and lens-modeling of 10,000 SDSS clusters \citep[see][]{Zitrin2011d}.

Due to the hierarchical growth of structure in the Universe, collapsed, virialized clusters should be found mostly at lower redshifts. These clusters make for excellent lenses as there is more mass concentrated in the cluster center, boosting the critical lensing area. According to this assumption, along with the dependency on the cosmological distances involved, lensing should be therefore optimized in clusters at redshifts of $z_{l}\sim0.2$. However, recent work has uncovered more large higher-redshift ($z_{l}\sim0.5$) lenses than expected by $\Lambda$CDM and related simulations, even after accounting for lensing bias  \citep[e.g.,][]{Zitrin2011a,Zitrin2011d,Meneghetti2011}.

The existence of high-redshift massive clusters at $z_{l}\ga1$ \citep{Rosati2009,Fassbender2011highzClus,Gobat2011,Planck2011highzClus,Santos2011highzmassCluster,Williamson2011SZclusters}, as well as the existence of evolved galaxies at high redshift, and other reported discrepancies such as the arc abundance and high concentrations, are also claimed to be unlikely given the predicted abundance of extreme perturbations of cluster sized masses in the standard $\Lambda$CDM scenario \citep[e.g.,][]{Daddi2007,Daddi2009,BroadhurstBarkana2008,Broadhurst2008,Jee2009,Jee2011,Richard2011,Zitrin2010,Zitrin2011a,Zitrin2011d,Zitrin2011c}. These claimed discrepancies possibly point towards a more extended early history of growth, or a non-Gaussian distribution of massive perturbations.

The galaxy cluster MACS J1206.2-0847 ($z$=0.4385; MACS1206 hereafter), is an X-ray selected system at
intermediate redshift found by the Massive Cluster Survey, MACS (\citealt{EbelingMacsCat2001,EbelingMacs12_2007,EbelingMacsFull2010}), and therefore constitutes an interesting lensing target. A first mass model for this cluster was presented by \citet{Ebeling2009}, based on 1-band HST/ACS imaging (F606W), combined with additional optical and NIR ground-based imaging. \citet{Ebeling2009} have identified one multiple system, consisting of a giant arc and its counter image at $z_{s}=1.036$, and presented a mass distribution for this cluster, though without constraining the profile due to the lack of sufficient high-resolution color-imaging, and correspondingly, other multiple-systems. The 16 HST bands chosen for the the Cluster Lensing And Supernova survey with Hubble (CLASH) project \citep{PostmanCLASHoverview}, ranging from the UV through the optical and to the IR, along with spectra from the VLT/VIMOS for some of the brighter arcs, enable us to obtain accurate redshifts for the multiply-lensed sources presented in this work. We use these data available to date, along with our well-tested approach to SL modeling \citep[e.g.,][]{Broadhurst2005a, Zitrin2009a, Zitrin2009b, Zitrin2010, Zitrin2011a,Zitrin2011b,Zitrin2011c}, to find a significant number of multiple images across the central field of MACS1206 so that its mass distribution and inner profile can be constrained for the first time, and with high precision.

The paper is organized as follows: In \S 2 we describe the
observations, and in \S 3 we detail the SL analysis. In \S 4 we report
and discuss the results. Throughout we adopt a concordance $\Lambda$CDM cosmology with
($\Omega_{\rm m0}=0.3$, $\Omega_{\Lambda 0}=0.7$, $h=0.7$). With these
parameters one arcsecond corresponds to a physical scale of 5.67 kpc
for this cluster (at $z=0.4385$; \citealt{Ebeling2009}). The reference center of our analysis
is fixed on the brightest cluster galaxy (BCG): RA = 12:06:12.15 Dec = -08:48:03.4
(J2000.0).

\begin{figure*}
 \begin{center}
  \includegraphics[width=160mm,trim=0mm 0mm 0mm 0mm,clip]{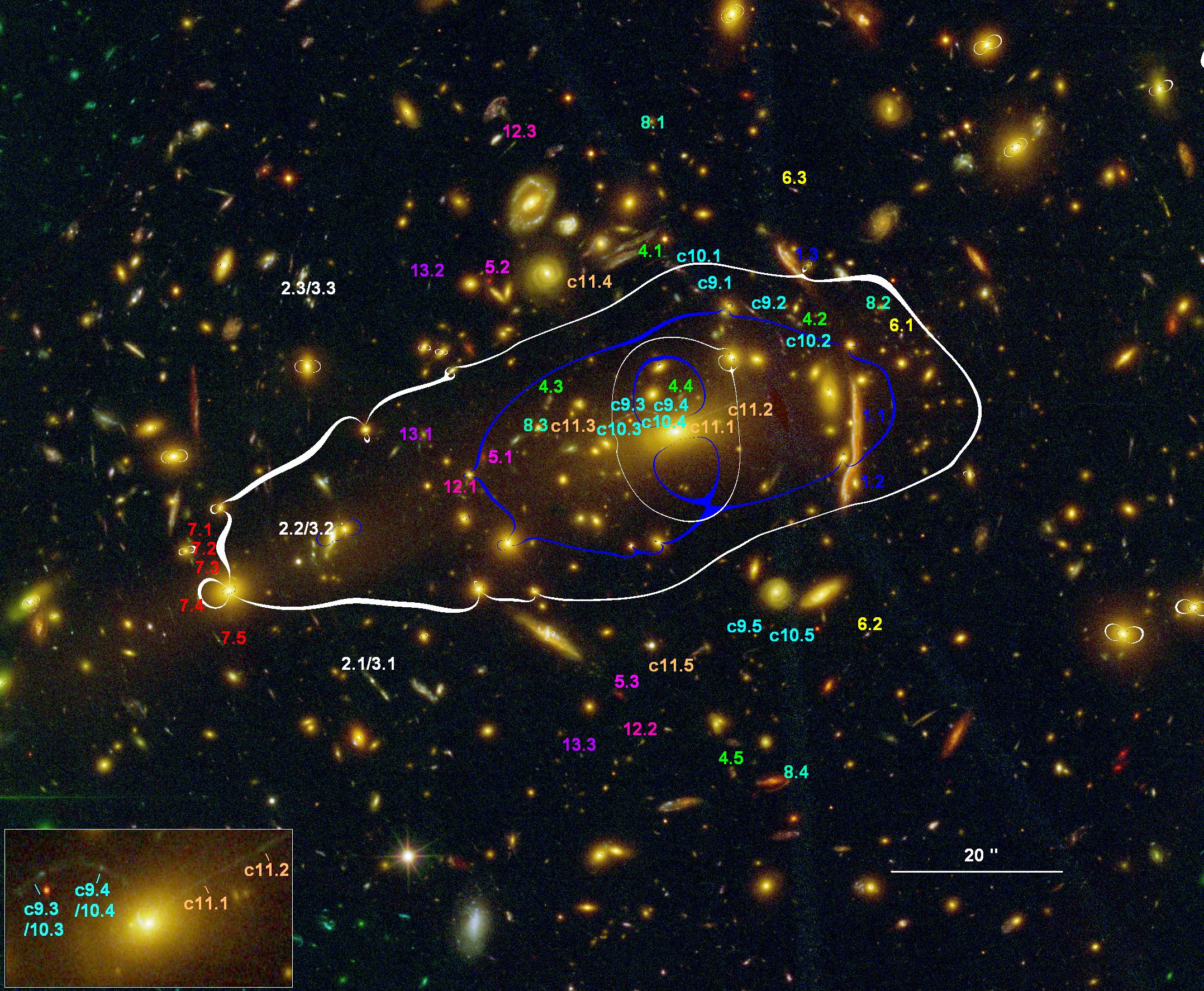}
 \end{center}
\caption{Galaxy cluster MACS1206 ($z=0.4385$) imaged
with HST/ACS/WFC3. North is up, East is left. We number the multiply-lensed images used and
uncovered in this work. The numbers indicate the 50 lensed images, 47 of which are uncovered here and correspond to (at least) 12 newly identified sources and candidates, and the different colors are used to distinguish between them. Note that candidate systems are marked in ``c''. Details on the each system are given in Table \ref{systems}. The overlaid white critical curve corresponds to system 4 at $z_{s}=2.54$, enclosing a critical area with an effective Einstein radius of $\simeq 160$ kpc at the redshift of this cluster ($28\arcsec$). Also plotted is a blue critical curve, which corresponds to the lower redshift of system 1, the giant arc system at $z_{s}=1.033$. The composition of this color image is Red=F105W+F110W+F125W+F140W+F160W, Green=F606W+F625W+F775W+F814W+F850LP, and Blue=F435W+F475W. The bottom-left inset shows an enlargement of the central core.}
\label{curves1206}
\end{figure*}

\section{Observations and Redshifts}\label{obs}

As part of the CLASH program, MACS1206 was observed with {\em HST} from 2011 March to 2011 July. This is the third of 25 clusters to be observed to a depth of 20 {\em HST} orbits in 16 filters with the Wide Field Camera 3 (WFC3) UVIS and IR cameras, and the Advanced Camera for Surveys (ACS) WFC. The images are processed for debias, flats, superflats, and darks, using standard techniques, and are then co-aligned and combined using drizzle algorithms to a scale of $0.065\arcsec /$ pixel. The full UVIS/ACS/WFC3-IR data set is then importantly used for multiple-images verification and measurement of their photometric redshifts using both the BPZ program \citep{Benitez2000, Benitez2004, Coe2006}, and LePhare (\citealt{ArnoutsLPZ1999,Ilbert2006BPZ}), where in practice 15 bands were used for the photometry, as observations for this cluster were still in progress during the preparation of this paper (nearly all 20 orbits had been completed for most filters. The F336W band was not used). Further details are presented in \citet{PostmanCLASHoverview}.

We obtain spectra for a number of multiple systems uncovered here, taken
as part of the VLT/VIMOS Large Programme 186.A-0798, which will
perform panoramic spectroscopy of 14 southern CLASH clusters,
targeting hundreds of cluster members per cluster and SL
features in their cores. Details on this program will be presented
elsewhere, when observations for one cluster are completed. For
each cluster, four VIMOS pointings are used, keeping one of the four
quadrants constantly locked on the cluster core, thus allowing long
exposures on the arcs, where exposure times for each pointing are about 45-60
minutes. By filling the inter-quadrant gaps, the final VIMOS layout covers
20-25\arcmin ~across. Either the low-resolution LR-Blue grism or the intermediate resolution orange MR grism is used, depending on the
photometric redshifts of the targets.

\begin{figure*}
 \begin{center}
   \includegraphics[width=180mm]{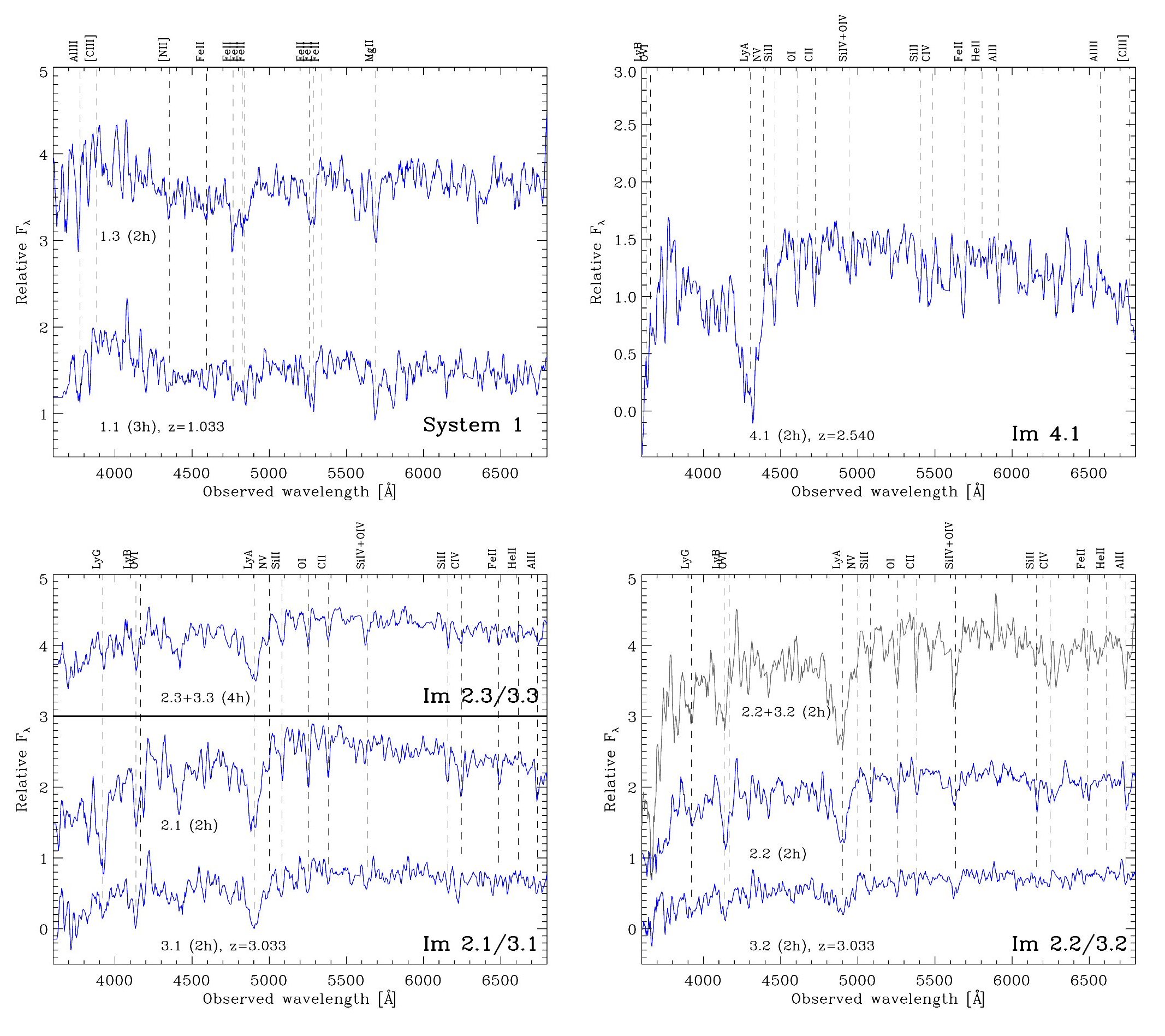}
 \end{center}
\caption{VLT/VIMOS spectra obtained for systems 1 to 4, at $z_{sys1}=1.033$, $z_{sys2,3}=3.033$, $z_{sys4}=2.540$. The exposure time for each spectrum is shown in
parentheses. For the $z=3.03$ systems, the pairs are blended into the
slits in some cases (2.2+3.2, 2.3+3.3) due to less ideal seeing conditions, though in some exposures the seeing was sufficient to resolve each individually.}
\label{spectraFig}
\end{figure*}

The spectra presented here are the results of the very first observations for this program, consisting
of four pointings with the LR-Blue grism, obtained in 2011 March-April, which yielded approximately
1000 redshifts. This configuration provides a spectral resolution of
$\sim 28$\AA~ with 1\arcsec~ slits and a useful wavelength coverage of
3700--6800\AA. Preliminary HST/CLASH data from the first two visits of MACS1206
were used to select images 1.[1,2,3], 2.[1,2,3], 3.[1,2,3], and 4.1,  as spectroscopic targets (see Figure \ref{curves1206}). The
slits ran along the NS direction. For some exposures the seeing was
very good so that separate spectra of the pairs 2.1/3.1 and 2.2/3.2 (separation of $\sim1\arcsec$)
were taken (and are shown in Figure \ref{spectraFig}, Table \ref{systems}), though in other cases the pairs were
blended in the slit. The spectrum of 1.2,
which contains the blend of the giant arc and a compact cluster
galaxy, is not shown. A spectrum of this arc, covering redder
wavelengths including the [OII] line, was however published in \citet{Ebeling2009}. Our spectroscopy also confirms that the four compact galaxies right on
the East and West side to the giant arc are early-type cluster members.

\section{Strong Lensing Modeling and Analysis}\label{model}

The approach to lens modeling we use here \citep[e.g.,][]{Broadhurst2005a, Zitrin2009b} begins with the
assumption that mass approximately traces light, so that the photometry of the red-sequence cluster member
galaxies is used as the starting point for our model. In particular, we use the F814W and F475W bands to filter in the brighter member galaxies ($m_{814}<23$ AB mag), and the F814W flux to derive the relative weight of each member. Using the extensive multi-band imaging and corresponding photometric redshifts, these galaxies can be then verified as members lying at the cluster redshift.

 We approximate the large scale distribution of cluster
mass by assigning a power-law mass profile to each
galaxy, the sum of which is then smoothed, using a 2D spline interpolation. The polynomial degree
of smoothing (S) and the index of the power-law
(q) are the most important free parameters determining
the mass profile: steeper power-laws and higher 2D polynomial degrees, generally entail a steeper profile \citep[see][]{Zitrin2009b}. A
 worthwhile improvement in fitting the location of the lensed images
 is generally found by expanding to first order the gravitational
 potential of this smooth component, equivalent to a coherent shear
 describing the overall matter ellipticity. The direction of the
 shear and its amplitude are free parameters, allowing for some flexibility in
 the relation between the distribution of dark matter (DM) and the distribution of
 galaxies, which cannot be expected to trace each other in detail. The
 total deflection field $\vec\alpha_T(\vec\theta)$, consists of the
 galaxy component, $\vec{\alpha}_{gal}(\vec\theta)$, scaled by a
 factor $K_{gal}$, the cluster DM component
 $\vec\alpha_{DM}(\vec\theta)$, scaled by (1-$K_{gal}$), and the
 external shear component $\vec\alpha_{ex}(\vec\theta)$:

\begin{equation}
\label{defTotAdd}
\vec\alpha_T(\vec\theta)= K_{gal} \vec{\alpha}_{gal}(\vec\theta)
+(1-K_{gal}) \vec\alpha_{DM}(\vec\theta)
+\vec\alpha_{ex}(\vec\theta).
\end{equation}

The best fit is
assessed by the minimum $\chi^{2}$ uncertainty in the image plane, and the errors are determined accordingly, by adopting a positional error of $2\arcsec$. This we have found is a typical value, following previous findings for the effect of large-scale structure (LSS) along the line-of-sight \citep{Jullo2010,Host2011LOS}, and by propagating a typical $\Delta z\sim0.1$ photo-$z$ uncertainty (for $z_{s}\sim2$) into the image-plane reproduction error. Note also, this image-plane minimization does not suffer from the bias involved with source-plane minimization, where solutions are biased by minimal scatter towards shallow mass profiles with correspondingly higher magnifications.

\begin{figure}
 \begin{center}
  \includegraphics[width=90mm]{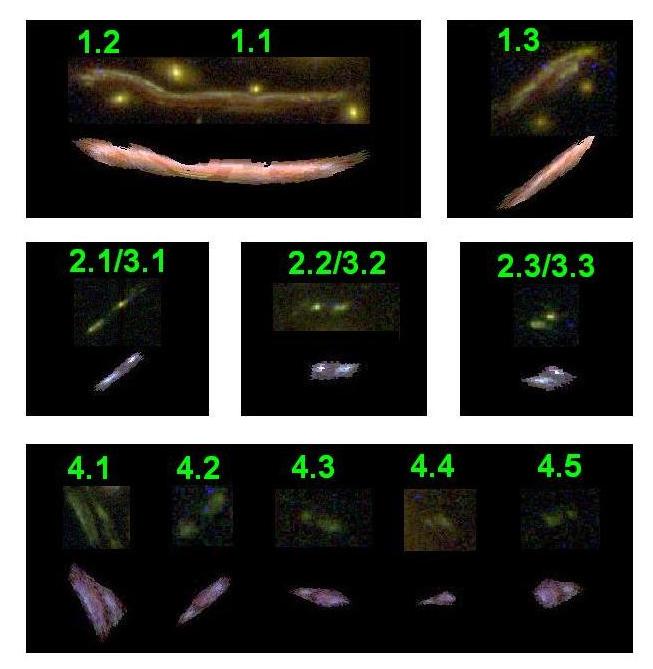}
 \end{center}
\caption{Reproduction of systems 1 to 4 by our model, compared with the real images. For each system we delens one image to the source plane and relens it back to the image-plane to obtain the other images of that system. More explicitly, we delens-relens, respectively, images 1.1, 2.2/3.2, and 4.1, each in the lensing-distance ratio expected from its redshift (1.033, 3.03, 2.54, respectively), to obtain the results shown here. The upper images in each row are the real images while the lower images are those reproduced by our model. As can be seen, our model reproduces the images of these systems with great accuracy. In these stamp images North is right, East is up, and some of the images are slightly zoomed-in for clarity. Note also, due to their similar position and symmetry, systems 2 and 3 were considered a single-system in the minimization of the mass model.}
\label{Rep1to4}
\end{figure}

It should be stressed that the multiple-images found here are
accurately reproduced by our model (e.g., Figure \ref{Rep1to4}), and are not simple identifications by
eye. Due to the small number of parameters in our model, it is initially well-constrained, enabling a reliable identification of other
multiple-images in the field. The mass model predictions are identified in the data and verified further by comparing the SEDs and photometric redshifts of the candidate multiple-images. The model is successively refined as additional sets of multiple
images are incorporated to improve the fit, importantly using also
their redshift information for better constraining the mass slope
through the cosmological relation of the $D_{ls}/D_{s}$ growth.

\begin{table*}
  \caption{Multiple-image systems and candidates}
\label{systems}
\begin{center}
\begin{tabular}{|c|c|c|c|c|c|c|c|c|c|}
\hline\hline
ARC & RA & DEC & BPZ $z_{phot}$&  LePhare $z_{phot}$& spec-$z$& $z_{model}$& $\Delta~Position$ &Comment\\
ID& (J2000.0)&(J2000.0)& (best) [95\% C.L.]& (best) [90\% C.L.]& & & (arcsecs) & \\
\hline
1.1 & 12:06:10.75 & -08:48:01.01 & -- [--]& -- [--]&1.033& (1.033)& 3.3& \\
1.2 & 12:06:10.82 & -08:48:08.95 & 1.04 [0.96--1.12]&1.06 [1.04--1.09] &(1.036)& " & 0.1& \citet{Ebeling2009}\\
1.3 & 12:06:11.29 & -08:47:43.44 & 1.01 [0.93--1.09]& 1.05 [1.03--1.06] &1.033& "& 3.3& \\
\hline
2.1 & 12:06:14.53 & -08:48:32.37& -- [--] &-- [--]&3.03& (3.03) & 0.2 & \\
2.2 & 12:06:15.00& -08:48:17.67& 3.40 [3.23--3.57]&3.19 [3.12--3.36]&3.03& "& 0.5 & \\
2.3 & 12:06:15.03& -08:47:48.07&3.68 [3.50--3.86]&3.64 [3.59--3.70]&3.03& "& 2.1 & \\
\hline
3.1 & 12:06:14.43&  -08:48:34.20& 3.73 [3.55--3.92]&3.65 [3.60--3.73]& 3.03& (3.03)& 0.2 & \\
3.2 &  12:06:15.00&  -08:48:16.50& -- [--] & -- [--]&3.03& "& 0.5& \\
3.3 &  12:06:15.01&  -08:47:48.65& 3.52 [3.34--3.70]& 3.62 [3.53--3.67]&3.03& "& 2.1& \\
\hline
4.1 &12:06:12.58 & -08:47:43.12 &2.99 [2.83--3.15]& 2.54 [2.48--2.60]& 2.54& (2.54)& 0& \\
4.2 & 12:06:11.22& -08:47:50.27&2.54 [2.36--2.71]&2.35 [2.18--2.64]&(2.54)& "& 0.7& \\
4.3 & 12:06:13.15&-08:47:59.84&2.35 [2.16--2.54]&1.93 [1.91--1.95]&(2.54)& "& 0.8& \\
4.4 & 12:06:12.119& -08:47:59.51&-- [--]&-- [--]&(2.54)& "& 1.5& \\
4.5 & 12:06:11.70& -08:48:43.04&3.04 [2.83--3.20]&2.67 [2.50--3.24]&(2.54)&"& 2.3& \\
\hline
5.1 &12:06:13.66 &-08:48:06.28 &1.87 [1.73--1.99]&1.76 [1.72--1.89]& --& $\simeq1.85$ & 0.1& \\
5.2 & 12:06:13.62& -08:47:45.85&1.80 [1.60--1.91]&1.63 [1.54--1.72]& --& "& 1.2& \\
5.3 &12:06:12.59 & -08:48:34.06&1.64 [1.46--1.84]&1.55 [1.09--1.74]&--&"& 0.9& \\
\hline
6.1 & 12:06:10.38 & -08:47:52.10&2.79 [2.64--2.94] &2.86 [2.77--2.93] &--& $\simeq2.8$ & 0& \\
6.2 & 12:06:10.64& -08:48:26.97&2.65 [2.29--2.79]& 2.53	[2.44--2.71]&--& "& 1.5& \\
6.3 & 12:06:11.21& -08:47:35.24&2.77 [2.62--2.92]& 2.53 [2.37--2.64]&--&"& 1.1& \\
\hline
7.1 & 12:06:15.98& -08:48:15.98&3.55 [3.37--3.91] & 3.47 [3.29--4.04]&--&  $\simeq3.7$& 0& affected by local galaxy weight\\
7.2 & 12:06:15.95& -08:48:17.21& -- [--]& -- [--]&--& "& 0& "\\
7.3 & 12:06:15.95& -08:48:18.45&4.02 [3.70--4.22]& 3.89 [3.34--4.07]&--&"& 0& "\\
7.4 & 12:06:15.90& -08:48:23.01& -- [--]&-- [--] &--& "& 0.7& "\\
7.5 & 12:06:15.75& -08:48:27.49&3.97 [3.78--4.17]& 3.68 [3.43--4.08]&--&"& 0.7& "\\
\hline
8.1 & 12:06:12.33& -08:47:28.61&5.44 [5.19--5.72] &5.52 [5.15--5.73]&--& $\simeq5.7$ & 2.0& \\
8.2 & 12:06:10.58& -08:47:49.49&5.42 [5.17--5.72]& 5.44 [5.10--5.75]&--& "& 3.2& \\
8.3 & 12:06:13.25& -08:48:03.81&5.42 [5.17--5.74]&5.42 [4.91--5.74]&--&"& 1.3& \\
8.4 & 12:06:11.36& -08:48:44.99&5.46 [5.21--5.71]& 5.51 [5.22--5.66] &--&"& 2.1& most probable\\
\hline
9.1 & 12:06:11.99& -08:47:46.89&1.74 [1.63--1.85] &1.74 [1.62--1.78]&--& $\simeq1$ & 2.7& candidate system\\
9.2 & 12:06:11.55& -08:47:49.36&1.76 [1.65--1.87]& 1.75 [1.67--1.79]&--& "& 1.7& " \\
9.3 & 12:06:12.53& -08:48:01.47& -- [--]& &--&"& 4.2& "\\
9.4 & 12:06:12.31& -08:48:01.21&-- [--] & &--&"& 1.1& "\\
9.5 & 12:06:11.48&-08:48:30.61&1.41 [0.85--1.78]& 1.08 [0.80-1.82]&--&"& 5.3& most probable\\
\hline
10.1 & 12:06:12.13& -08:47:44.48&1.68 [1.48--1.79] & 1.54 [1.50--1.65]&--&  $\simeq1.4$ & 5.7& candidate system\\
10.2 & 12:06:11.32& -08:47:52.81&1.57 [1.11--1.72]& 1.60 [1.02--1.76]&--& "& 3.5& "\\
10.3 & 12:06:12.53& -08:48:01.47&-- [--]& &--&"& 3.1& degeneracy with 9.3\\
10.4 & 12:06:12.31& -08:48:01.21&-- [--]& &--&"& 0& degeneracy with 9.4\\
10.5 & 12:06:11.31&-08:48:32.83 &1.38 [0.8--1.86]& 0.94	[0.90--1.07]&--&"& 1.6& most probable\\
\hline
11.1 & 12:06:12.04& -08:48:02.25&-- [--]  & &--& $\simeq1.3$ & 6.4& candidate system\\
11.2 & 12:06:11.77& -08:48:00.49&-- [--] & &--& "& 2.9& "\\
11.3 & 12:06:13.08& -08:48:04.00&1.42 [1.13--1.78]&1.72 [1.72--1.72] &--&"& 1.9& "\\
11.4 & 12:06:12.88& -08:47:44.75&1.12 [0.31--1.48]&1.10 [0.04--1.15] &--&"& 0& ", bimodal\\
11.5 & 12:06:11.99& -08:48:31.98&1.31 [1.19--1.76]& 1.32 [1.07--1.80]&--&"& 2.6& "\\
\hline
12.1 & 12:06:13.82& -08:48:11.03&3.78 [3.50--4.05] & 3.63 [3.26--4.01]&--& $\simeq4$ & 3.2& \\
12.2 & 12:06:12.42& -08:48:39.53&3.91 [3.44--4.13]& 0.28 [0.11--3.43]&--& "& 0.1& bimodal\\
12.3 & 12:06:13.43& -08:47:29.91&3.97 [3.70--4.17]& 3.88 [3.61--4.06]&--&"& 2.7& \\
\hline
13.1 & 12:06:14.16& -08:48:05.11&2.88 [2.65--3.12] &2.90 [2.32--3.16] &--& $\simeq3.6$ &2.4 & \\
13.2 & 12:06:14.09& -08:47:46.37&3.64 [3.30--3.82]&3.43 [0.23--3.71] &--& "& 0& bimodal\\
13.3 & 12:06:12.89& -08:48:41.09&3.15 [0.2--3.41]& 2.97	[0.08--3.43] &--&"& 0.8& " \\
\hline\hline
\end{tabular}
\tablecomments{Multiple-image systems and candidates used and uncovered by our model. Columns are: arc ID; RA
    and DEC in J2000.0; best photo-$z$ using BPZ \citep{Benitez2000, Benitez2004, Coe2006}, along with its 95\% confidence level; best photo-$z$ using LePhare \citep{ArnoutsLPZ1999,Ilbert2006BPZ}, along with its 90\% confidence level; spectroscopic redshift, spec-$z$ (images marked in parenthesis were not spectroscopically measured here); $z_{model}$, estimated redshift for the arcs which lack spectroscopy as predicted by the mass model; $\Delta~Position$, difference in arcseconds between the images reproduced by the model and the observed images; comments. System 1 was uncovered by \citet{Ebeling2009} who measured the redshift of images 1.1 and 1.2 spectroscopically at $z=1.036$; our spectroscopy yields $z=1.033$ for this system. Note that we interpret the long arc (images 1.1/1.2) as a double-lensed image, though \citet{Ebeling2009} identify it to consist of several (partially) lensed images. Either interpretation has only a very local (negligible) effect on the model. All other systems listed above are found in our work, where we obtain VLT/VIMOS spectroscopy for systems 1 to 4. Note that unusually large errors in the photo-$z$ imply a bimodal distribution, such cases are specified in the comments. We denote in the comments column where the most probable image was chosen but other candidates are seen nearby. Systems 9 to 11 are candidate systems, since only some of the ($<1\sigma$) models can reproduce them, or simply since their photo-$z$ disagrees with the model prediction. Also, systems 9 and 10 show similar symmetry to that of system 4, strengthening their identification on one hand, but only 2 radial images are seen. Due to the BCG light it is hard to determine unambiguously if, or to which, of these two systems they belong. Note also, that if images 11.1 and 11.2 are not multiple images, but, say, a jet coming out of the BCG (see Figure \ref{curves1206}), then images 11.3 to 11.5 may constitute an individual system. All other systems we consider as secure identifications, in the context of the photometric redshifts, internal details, and the reproduction by our model.}
\end{center}
\end{table*}

\begin{figure}
 \begin{center}
  \includegraphics[width=90mm]{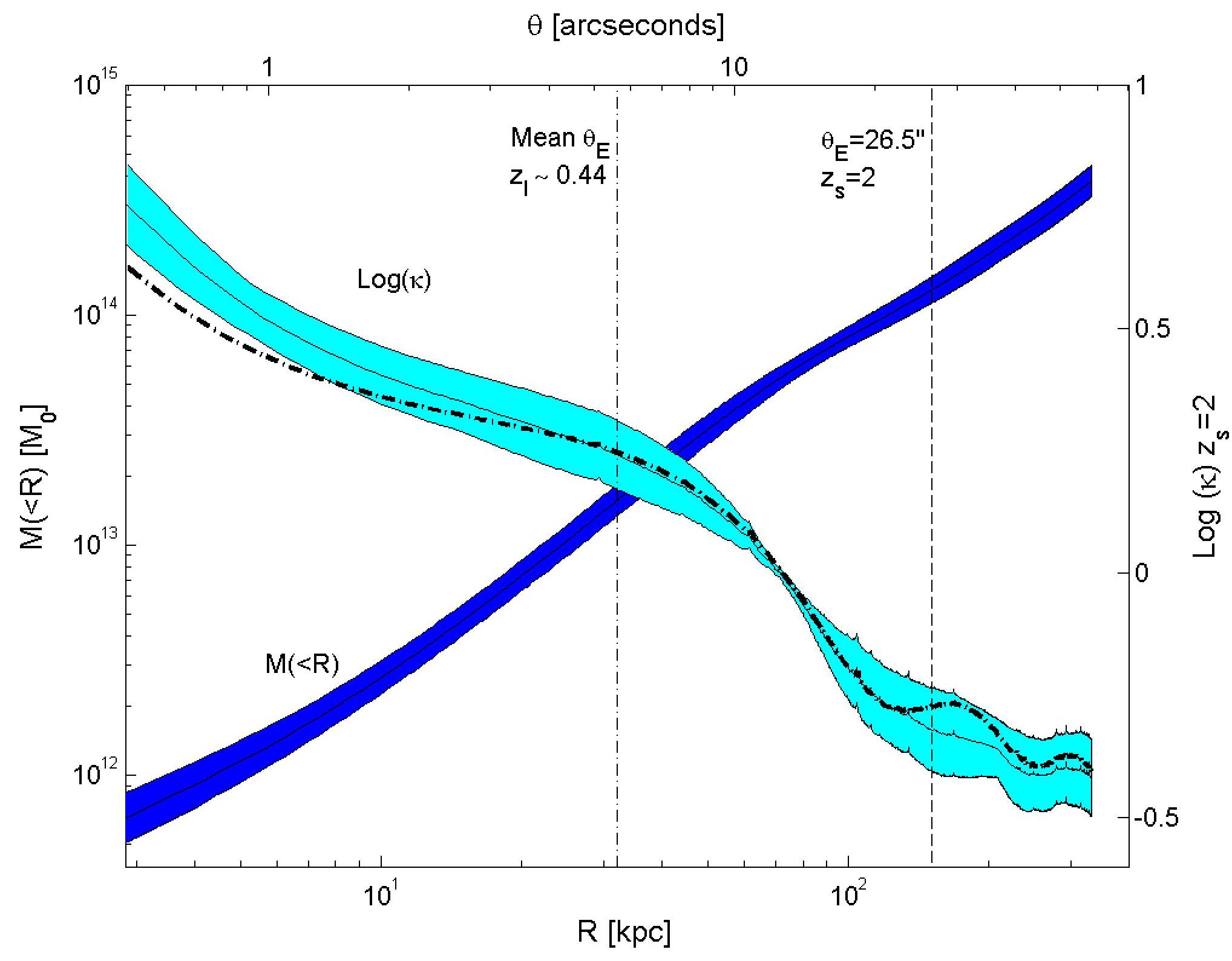}
 \end{center}
\caption{Projected total mass profile. The \emph{light blue} curve shows the radial surface mass-density profile in units of the critical surface density ($\kappa$; right-side $y$-axis), for a source redshift of $z_s=2$. The \emph{dark blue} curve shows the overall enclosed mass per radius, $M(<R)$ (left-side $y$-axis). The widths of the blue curves indicate the $\sim1\sigma$ errors. The thick dash-dotted curve is the (preliminary) best-fit model resulting from a $4\times10^4$ step Monte-Carlo Markov Chain (MCMC) with Metropolis-Hastings
algorithm, which allows also the BCG mass to vary. The results are very similar, with only some discrepancy at radii below $\sim1\arcsec$. The details of this MCMC method will be presented elsewhere. The dash-dotted vertical line denotes the mean Einstein radius ($z_{s}=2$) distribution from $3000$ SDSS clusters at $z\sim0.44$ \citep{Zitrin2011d}, and the vertical dashed line denotes the Einstein radius of MACS1206 (see \S 4 for this comparison).}
\label{profileAdi}
\end{figure}

\section{Results and Conclusions}\label{results}

Using CLASH imaging, we have identified 47 new multiple-images in MACS1206, corresponding to 12 distant sources. These images, along with the previously known arc \citep{Ebeling2009}, span the redshift range $1\la z\la5.5$ and thus allow us to derive a robust mass distribution and constrain the profile of this cluster, for the first time.

We made use of the location and redshift of all secure multiple-images (i.e., excluding candidate systems 9-11) to fully constrain the mass model and profile. We obtained VLT/VIMOS spectra for systems 1 to 4, at redshifts 1.033, 3.03, 3.03, and 2.54, respectively, to help pin down the profile with greater accuracy. Details on the multiple systems are found in Figure \ref{curves1206} and Table \ref{systems}, where in Figure \ref{Rep1to4} we show examples for reproductions of some of the multiple-systems by our model. The $\chi^{2}$ for the best model is 25.7. With 35 secure multiply-lensed images and 6 free parameters, this yields a reduced $\chi^{2}$ of $\sim1$. Estimating the accuracy of our mass distribution quantitatively, for all secure images we obtain an average image-plane reproduction uncertainty of $1.3\arcsec$ per image, with an image-plane $rms$ of $1.8\arcsec$, typical to parametric mass models for clusters with many multiple images \citep{Broadhurst2005a,Halkola2006,LimousinOnA1689,Zitrin2009b}. Also, this \emph{rms} value is realistic given the $\sim1\arcsec$ noise level expected from LSS along the line-of-sight \citep[e.g.][]{Jullo2010,DAloisioNatarajan2011LSS}, the many multiple images used, and the small number of free parameters in our modeling.

For a source at $z_s=2.54$, the critical curves enclose a relatively large area, with an effective Einstein
radius of $r_{E}=28\pm3\arcsec$, or $\simeq$158 kpc at the redshift
of the cluster. A projected mass of $1.34\pm0.15 \times
10^{14}M_{\odot}$ is enclosed by this critical curve for this source redshift (see Figure
\ref{curves1206}). For the lower source redshift of system 1, $z_s=1.033$, the Einstein radius is $\simeq17\arcsec$, enclosing a projected mass of $0.8\pm0.1 \times10^{14}M_{\odot}$. For comparison, our model encloses within 21\arcsec, a projected mass of $M(<21\arcsec) \simeq 1\pm0.1 \times10^{14} M_{\odot}$, while \citet{Ebeling2009} found similarly, although based on only one system, a projected mass of $M(<21\arcsec) = 1.12\pm0.05 \times10^{14} M_{\odot}$, consistent with our result. In addition, as a consistency check for the very inner profile around the BCG, we
compared our result to the F160W light. By estimating the
stellar mass profile from the F160W surface brightness photometry,
assuming a Kroupa IMF, the stellar mass is approximately 10\% of the
total mass within 30 kpc, which is approximately half the BCG
effective radius. We defer to forthcoming papers the exact assessment of the DM mass distribution steepness in the very inner region (say, below $\sim5$ kpc) after taking into account the stellar and gas baryonic contributions.

The corresponding critical curves for different redshifts are plotted on the cluster image in
Figure \ref{curves1206}, along with the multiply-lensed systems. The resulting total mass profile is shown in Figure \ref{profileAdi}, for which we measure a slope of $d\log \Sigma/d\log r\simeq -0.55\pm 0.1$ (in the range [1\arcsec,53\arcsec], or $5\la r \la300$ kpc; about twice the Einstein radius), similar to other usually-relaxed and well-concentrated lensing clusters \citep{Broadhurst2005a,Zitrin2009b,Zitrin2010}. However, it is not perfectly clear whether MACS1206 is indeed a relaxed cluster; X-ray and optical light contours \citep[see e.g.,][]{Ebeling2009} indeed show an approximately (circularly) symmetric distribution, without a prominent sign of recent merger. However, a high velocity dispersion (1580 km s$^{-1}$; \citealt{GilmourUnRelaxed2009}), along with the excessive X-ray luminosity ($2.4 \times 10^{45}$ erg s$^{-1}$, [0.1-–2.4] keV) and temperature ($11.6\pm0.7$ keV; see \citealt{Ebeling2009}), may imply a merger along the line of sight (see also \citealt{PostmanCLASHoverview}). A full assessment of the degree of relaxation of this system will be soon enabled, by the dynamical
analysis from several hundreds member velocities we
are currently collecting in our spectroscopic program, as
well as from the combination of other mass diagnostics.

We note that given its redshift, MACS1206 has a relatively large Einstein radius. Previous studies have shown that other MACS clusters at a redshift of $z\sim0.5$ distribute around this Einstein radius size ($\theta_E\simeq28\arcsec$), but with a noticeable discrepancy from expectations of the $\Lambda$CDM model and related simulations \citep[e.g.,][]{Zitrin2011a,Meneghetti2011}, even after taking into account triaxiality-induced lensing bias.

Recently we have applied our lens-modeling technique to an unprecedentedly large sample of 10,000 SDSS clusters, to deduce a representative distribution of Einstein radii \citep[see][]{Zitrin2011d}, covering the full cluster mass-range. For the redshift bin corresponding to MACS1206 (see Figure 5 therein), the distribution for $z_{s}=2$ peaks below 10\arcsec~(with median and mean $\theta_{E}$ of 4.1\arcsec and 5.6\arcsec, respectively), and rapidly declines towards larger radii. As may be expected from the different selection criteria, the Einstein radius of the X-ray selected cluster MACS1206 ($\theta_{E}\simeq26.5\arcsec$ for $z_{s}=2$) sits at the (far) tail of the distribution: only $\simeq1.3\%$ of the optically-selected SDSS clusters examined at this redshift have similar (or larger) Einstein radii, and as much corresponding enclosed mass, as found in MACS1206. The lensing model for MACS1206 presented here, enabled by deep, multi-band HST imaging, constitutes an important example of the inner mass distributions of systems lying at the high-end of the cluster mass function.

\section*{acknowledgments}

We thank the anonymous reviewer of this manuscript for useful comments. The CLASH Multi-Cycle Treasury Program (GO-12065) is based on observations made with the NASA/ESA Hubble Space Telescope. The Space Telescope Science Institute is operated by the Association of Universities for Research in Astronomy, Inc. under NASA contract NAS 5-26555. Part of this work is based on data collected at the Subaru Telescope, which is operated by the National Astronomical Society of Japan and on data collected at the Very Large Telescope at the ESO Paranal Observatory, under Programme ID 186.A-0798.
\bibliographystyle{apj}
\bibliography{outDan3marc}

\end{document}